\documentclass[english]{article}
\usepackage[T1]{fontenc}
\usepackage[latin1]{inputenc}
\usepackage{geometry}
\usepackage{amsmath}
\usepackage{setspace}
\usepackage{amssymb}

\makeatletter

\providecommand{\tabularnewline}{\\}

\newcommand{\lyxaddress}[1]{
\par {\raggedright #1
\vspace{1.4em}
\noindent\par}
}

\usepackage{babel}
\makeatother
\begin{document}

\title{Unified Split Octonion Formulation of Dyons}

\author{P. S. Bisht$^{\text{(1)}}$,~ Shalini Dangwal$^{\text{(1)}}$~
and~ O.P.S. Negi$^{\text{(2)}}$%
\thanks{Permanent Address- Department of Physics, Kumaun University, S. S.
J. Campus, Almora -263601 (U.A.) INDIA%
}}

\maketitle
\begin{singlespace}

\lyxaddress{\begin{center}
$^{\text{(1)}}$Department of Physics\\
Kumaun University\\
S. S. J. Campus\\
 Almora -263601 (U.K.) India
\par\end{center}}
\end{singlespace}

\lyxaddress{\begin{center}
$^{\text{(2)}}$Institute of Theoretical Physics\\
 Chinese Academy of Sciences\\
 KITP Building Room No.- 6304\\
Hai Dian Qu Zhong Guan Chun Dong Lu\\
 55 Hao , Beijing - 100080, P.R.China 
\par\end{center}}

\lyxaddress{\begin{center}
Email:- ps\_bisht123@rediffmail.com\\
shalini\_dangwal@rediffmail.com\\
ops\_negi@yahoo.co.in
\par\end{center}}

\begin{abstract}
Demonstrating the split octonion formalism for unified fields of dyons
(electromagnetic fields) and gravito-dyons (gravito-Heavisidian fields
of linear gravity), relevant field equations are derived in compact,
simpler and manifestly covariant forms. It has been shown that this
unified model reproduces the dynamics of structure of fields associated
with individual charges (masses) in the absence of others.
\end{abstract}

\section{Introduction}

~~~~~~~~Magnetic monopoles \cite{key-1} were advocated to
symmetrize Maxwell's equations in a manifest way that the mere existence
of an isolated magnetic charge implies the quantization of electric
charge and accordingly the considerable literature \cite{key-2,key-3,key-4,key-5,key-6,key-7}
has come in force. The fresh interests are enhanced with the idea
of t' Hooft \cite{key-8} and Polyakov \cite{key-9} that the classical
solutions having the properties of magnetic monopoles may be found
in Yang - Mills gauge theories. Julia and Zee \cite{key-10} extended
it to construct the theory of non Abelian dyons (particles \cite{key-2,key-3}
carrying simultaneously electric and magnetic charges). In view of
the explanation of CP-violation in terms of non-zero vacuum angle
of world \cite{key-11}, the monopoles are necessary dyons and Dirac
quantization condition permits dyons to have analogous electric charge.The
quantum mechanical excitation of fundamental monopoles include dyons
which are automatically arisen \cite{key-5,key-7} from the semi-classical
quantization of global charge rotation degree of freedom of monopoles.
Accordingly, the self-consistent and manifestly covariant theory of
generalized electromagnetic fields associated with dyons (particles
carrying electric and magnetic charges) has been discussed by us \cite{key-12,key-13}. 

The close analogy between Newton's gravitation law and Coulomb's law
of electricity led many authors to investigate further similarities,
such as the possibility that the motion of mass-charge could generate
the analogous of a magnetic field which is produced by the motion
of electric-charge, i.e. the electric current. So, there should be
the mass current would produce a magnetic type field namely 'gravitomagnetic'
field. Maxwell \cite{key-14} in one of his fundamental works on electromagnetism,
turned his attention to the possibility of formulating the theory
of gravitation in a form corresponding to the electromagnetic equations.
In 1893 Heaviside \cite{key-15} investigated the analogy between
gravitation and electromagnetism where he explained the propagation
of energy in a gravitational field, in terms of a gravito electromagnetic
Poynting vector, even though he (just as Maxwell did) considered the
nature of gravitational energy a mystery. The analogy has also been
explored by Einstein \cite{key-16}, in the framework of General Relativity,
and then by Thirring \cite{key-17} and Lense and Thirring \cite{key-18},
that a rotating mass generates a gravito magnetic field causing a
precession of planetary orbits. Expounding the basics of the gravito
electromagnetic form of the Einstein equations,theory of gravito magnetism
has also been reviewed by Ruggiero-Tartaglia \cite{key-19}. 

As such, this analogy describes a structural symmetry between linear
gravitational and usual electromagnetic fields and leads the asymmetry
in Einstein's linear equation of gravity and accordingly suggests
the existence \cite{key-20,key-21} of gravitational analogue of magnetic
monopole. Like magnetic field, Cantani \cite{key-22} introduced a
new field (i.e. namely the Heavisidian field) depending upon the velocities
of gravitational charges (masses) and derived the covariant equations
(Maxwell's equations) of linear gravitational fields. Avoiding the
use of arbitrary string variables \cite{key-1}, we \cite{key-23,key-24}
have also formulated manifestly covariant theory of gravito-dyons
in terms of two four-potentials and maintained the structural symmetry
between generalized electromagnetic fields of dyons and generalized
gravito-Heavisidian fields of gravito-dyons.

There has been a revival in the formulation of natural laws so that
there exists \cite{key-25} four-division algebras consisting the
algebra of real numbers ($\mathbb{R}$), complex numbers ($\mathbb{C}$),
quaternions ($\mathbb{H}$) and Octonion ($\mathcal{O}$). All four
algebra's are alternative with totally anti symmetric associators.
Quaternions \cite{key-26} were very first example of hyper complex
numbers have been widely used \cite{key-27} to the various applications
of mathematics and physics. Since octonions share with complex numbers
and quaternions, many attractive mathematical properties, one might
except that they would be equally as useful as others. Octonion \cite{key-28}
analysis has been widely discussed by Baez \cite{key-29}. It has
now played an important role in the context of various physical problems
\cite{key-30} of higher dimensional supersymmetry, super gravity
and super strings etc. In recent years, it has also drawn interests
of many \cite{key-31} towards the developments of wave equation and
octonion form of Maxwell's equations. We have also studied \cite{key-32}
octonion electrodynamics, dyonic field equation and octonion gauge
analyticity of dyons consistently and obtained the corresponding field
equations (Maxwell's equations) and equation of motion in compact
and simpler formulation. Keeping these applications of octonions in
mind, in the present paper, we have applied the formalism of split
octonions to develop an unified model for generalized electromagnetic
fields of dyons and those for generalized Gravito-Heavisidian fields
of gravito dyons with the simultaneous existence of electric, magnetic,
gravitational and Heavisidian charges (masses). We have thus obtained
manifestly covariant forms of relevant field equations and derived
the corresponding quantization parameters in consistent, compact,
simpler forms. It has been shown that this unified theory reproduces
the dynamics of individual charges (masses) in the absence of others.

\section{Octonion Definition}

~~~~~~~An octonion $x$ is expressed as a set of eight real
numbers

\begin{eqnarray}
x=(x_{0},\, x_{1},....,\, x_{7}) & = & x_{0}e_{0}+x_{1}e_{1}+x_{2}e_{2}+x_{3}e_{3}+x_{4}e_{4}+x_{5}e_{5}+x_{6}e_{6}+x_{7}e_{7}\nonumber \\
= & x_{0}e_{0} & +\sum_{A=1}^{7}x_{A}e_{A}\,\,\,\,\,\,\,\,\,\,\,(A=1,2,.....,7)\label{eq:1}\end{eqnarray}
where $e_{A}(A=1,2,.....,7)$ are imaginary octonion units and $e_{0}$is
the multiplicative unit element. Set of octets $(e_{0},e_{1},e_{2},e_{3},e_{4},e_{5},e_{6},e_{7})$
are known as the octonion basis elements and thus satisfy the following
multiplication rules

\begin{eqnarray}
e_{0}=1,\,\,\,\,\,\,\: & e_{0}e_{A}=e_{A}e_{0}=e_{A}\,\,\,\,\,\,\,\,\, & (A=1,2,......7)\nonumber \\
e_{A}e_{B}= & -\delta_{AB}e_{0}+f_{ABC}\, e_{C}.\,\,\,\,\,\,\,\,\,\, & (A,B,C=1,2,......7)\label{eq:2}\end{eqnarray}
The structure constants $f_{ABC}$ are described as completely antisymmetric
and take the value $1$ for following combinations\cite{key-29,key-30,key-31};

\begin{eqnarray}
f_{ABC}= & +1 & =(123),\,\,(471),\,\,(257),\,\,(165),\,\,(624),\,\,(543),\,\,(736).\label{eq:3}\end{eqnarray}

\begin{center}
It is to be noted that the summation convention is used for repeated
indices. Here the octonion algebra $\mathcal{O}$ is described over
the algebra of rational numbers having the vector space of dimension
$8$. Octonion algebra is non associative and multiplication rules
for its basis elements given by equations (\ref{eq:2},\ref{eq:3})
are then generalized in the following table:
\par\end{center}

\begin{tabular}{|c||c||c||c||c||c||c||c|}
\hline 
$\cdot$&
$e_{1}$&
$e_{2}$&
$e_{3}$&
$e_{4}$&
$e_{5}$&
$e_{6}$&
$e_{7}$\tabularnewline
\hline
\hline 
$e_{1}$&
$-1$&
$e_{3}$&
$-e_{2}$&
$e_{7}$&
$-e_{6}$&
$e_{5}$&
$-e_{4}$\tabularnewline
\hline
\hline 
$e_{2}$&
$-e_{3}$&
$-1$&
$e_{1}$&
$e_{6}$&
$e_{7}$&
$-e_{4}$&
$-e_{5}$\tabularnewline
\hline
\hline 
$e_{3}$&
$e_{2}$&
$-e_{1}$&
$-1$&
$-e_{5}$&
$e_{4}$&
$e_{7}$&
$-e_{6}$\tabularnewline
\hline
\hline 
$e_{4}$&
$-e_{7}$&
$-e_{6}$&
$e_{5}$&
$-1$&
$-e_{3}$&
$e_{2}$&
$e_{1}$\tabularnewline
\hline
\hline 
$e_{5}$&
$e_{6}$&
$-e_{7}$&
$-e_{4}$&
$e_{3}$&
$-1$&
$-e_{1}$&
$e_{2}$\tabularnewline
\hline
\hline 
$e_{6}$&
$-e_{5}$&
$e_{4}$&
$-e_{7}$&
$-e_{2}$&
$e_{1}$&
$-1$&
$e_{3}$\tabularnewline
\hline
\hline 
$e_{7}$&
$e_{4}$&
$e_{5}$&
$e_{6}$&
$-e_{1}$&
$-e_{2}$&
$-e_{3}$&
$-1$\tabularnewline
\hline
\end{tabular}

\begin{description}
\item [{~~~~~~~~~~~~~~Table1-}] Octonion Multiplication table
\end{description}
Hence we have

\begin{eqnarray}
e_{A}(e_{B}e_{C}) & \neq & (e_{A}e_{B})e_{C}\label{eq:4}\end{eqnarray}
and the commutation rules for octonion basis elements are given by

\begin{eqnarray}
\left[e_{A},\,\, e_{B}\right] & = & 2f_{ABC}e_{C};\nonumber \\
\left\{ e_{A},\,\, e_{B}\right\}  & = & -\delta_{AB}e_{0};\label{eq:5}\end{eqnarray}
where brackets $[\,\,]$ and $\{\,\,\}$ are used respectively for
commutation and the anti commutation relations while $\delta_{AB}$
is the usual Kroneckar delta-Dirac symbol.Octonion conjugate is thus
defined as,

\begin{eqnarray}
\bar{x} & = & x_{0}e_{0}-x_{1}e_{1}-x_{2}e_{2}-x_{3}e_{3}-x_{4}e_{4}-x_{5}e_{5}-x_{6}e_{6}-x_{7}e_{7}\nonumber \\
= & x_{0}e_{0} & -\sum_{A=1}^{7}x_{A}e_{A}\,\,\,\,\,\,\,\,\,\,\,(A=1,2,.....,7)\label{eq:6}\end{eqnarray}
where we have used the conjugates of basis elements as

\begin{eqnarray}
\overline{e_{0}}=e_{0};\,\,\,\, & \overline{e_{A}} & =-e_{A}.\label{eq:7}\end{eqnarray}
An Octonion can be decomposed in terms of its scalar $(Sc(x))$ and
vector $(Vec(x))$ parts as 

\begin{eqnarray}
Sc(x) & = & \frac{1}{2}(x+\bar{x})=x_{0}\nonumber \\
Vec(x) & = & \frac{1}{2}(x-\bar{x})=\sum_{A=1}^{7}x_{A}e_{A}\label{eq:8}\end{eqnarray}
Conjugates of product of two octonions and its own are described as

\begin{eqnarray}
(\overline{xy}) & = & \overline{y}\,\overline{x}\,\,\,;\,\,\,\,\,\,\,\overline{(\bar{x})}\,\,=x\label{eq:9}\end{eqnarray}
while the scalar product of two octonions is defined as 

\begin{eqnarray}
\left\langle x\,,\, y\,\right\rangle  & =\sum_{\alpha=0}^{7} & x_{\alpha}y_{\alpha}=\frac{1}{2}(x\,\bar{y}+y\,\bar{x})=\frac{1}{2}(\bar{x}\, y+\bar{y}\, x)\label{eq:10}\end{eqnarray}
which can be written in terms of octonion units as

\begin{eqnarray}
\left\langle e_{A}\,,\, e_{B}\,\right\rangle  & = & \frac{1}{2}(e_{A}\overline{e_{B}}+e_{B}\overline{e_{A}})=\frac{1}{2}(\overline{e_{A}}e_{B}+\overline{e_{B}}e_{A})=\delta_{AB}.\label{eq:11}\end{eqnarray}
Following Catto \cite{key-30}, let us define 

\begin{eqnarray}
e_{AB} & = & \frac{1}{2}(\overline{e_{A}}e_{B}-\overline{e_{B}}e_{A})\label{eq:12}\end{eqnarray}
and \begin{eqnarray}
e'_{AB} & = & \frac{1}{2}(e_{A}\overline{e_{B}}-e_{B}\overline{e_{A}}).\label{eq:13}\end{eqnarray}
Hence we may write

\begin{eqnarray}
\overline{e_{A}}e_{B} & =\frac{1}{2}(\overline{e_{A}}e_{B}+\overline{e_{B}}e_{A})+ & \frac{1}{2}(\overline{e_{A}}e_{B}-\overline{e_{B}}e_{A})=\delta_{AB}+e_{AB}\label{eq:14}\end{eqnarray}
and

\begin{eqnarray}
e_{A}\overline{e_{B}} & =\frac{1}{2}(e_{A}\overline{e_{B}}+e_{B}\overline{e_{A}})+ & \frac{1}{2}(e_{A}\overline{e_{B}}-e_{B}\overline{e_{A}})=\delta_{AB}+e'_{AB}.\label{eq:15}\end{eqnarray}
Equations (\ref{eq:12}) and (\ref{eq:13}) may be interpreted as
the dyadic anti symmetric tensors and can be written component wise
as 

\begin{eqnarray}
e_{AB} & = & e'_{AB}=-f_{ABC}e_{C};\,\,\,\,\,\,\,\,\,\,\,\, e_{0A}=e'_{0A}=e_{A}.\label{eq:16}\end{eqnarray}
It shows that octonions describe the covariant formulations in eight
dimensional space. The norm of the octonion $N(x)$ is defined as

\begin{eqnarray}
N(x)=\overline{x}x & =x\,\bar{x} & =\sum_{\alpha=0}^{7}x_{\alpha}^{2}e_{0}\label{eq:17}\end{eqnarray}
which is zero if $x=0$, and is always positive otherwise. It also
satisfies the following property of normed algebra

\begin{eqnarray}
N(xy) & =N(x)N(y) & =N(y)N(x).\label{eq:18}\end{eqnarray}
As such, for a nonzero octonion $x$ , we define its inverse as

\begin{eqnarray}
x^{-1} & = & \frac{\bar{x}}{N(x)}\label{eq:19}\end{eqnarray}
which shows that

\begin{eqnarray}
x^{-1}x & =xx^{-1} & =1.e_{0}\nonumber \\
(xy)^{-1} & = & y^{-1}x^{-1}.\label{eq:20}\end{eqnarray}
Equation (\ref{eq:4}) shows that octonions are not associative and
thus do not form the group in their usual form. Non-associativity
of octonion algebra $\mathcal{O}$ is provided by the associator \cite{key-30,key-31,key-32,key-33}defined
for any $3$ octonions as follows,

\begin{eqnarray}
(x,y,z) & = & (xy)z-x(yz)\,\,\,\,\,\forall x,y,z\in\mathcal{O}\label{eq:21}\end{eqnarray}
which gives rise to the associator for octonion units as 

\begin{eqnarray}
(e_{A},e_{B},e_{C}) & = & 2\,\varepsilon_{ABCD}e_{D}\,\,\,\,\,.\forall\,\,(A,B,C,D=1,2,...,7).\label{eq:22}\end{eqnarray}
Here $\varepsilon_{ABCD}$ are totally antisymmetric and equal to
unity for the following $7$combinations ,

\begin{eqnarray}
1247,1265, & 2345,\,\,2376, & \,\,3146,\,\,3157\,\, and\,\,4576.\label{eq:23}\end{eqnarray}
On the other hand, the quaternion algebra $\mathbb{H}$ satisfies
the associativity and forms a group under multiplication. It is described
as the sub algebra of octonions and thus can be represented in terms
of unit matrix $\hat{1}$ and Pauli matrices $\sigma_{j}$as 

\begin{eqnarray}
e_{0}\rightarrow\sigma_{0}=\hat{1}\,,\,\,\,\mbox{and} & e_{j}\rightarrow & -i\sigma_{j}\,\,\,(\forall\,\,\, j=1,2,3)((i=\sqrt{-1})).\label{eq:24}\end{eqnarray}
It is trivial to check that the above map is an isomorphism i.e.

\begin{eqnarray}
e_{j}e_{k}\Rightarrow-\sigma_{j}\sigma_{k}=-(\delta_{jk}+i\,\varepsilon_{jkl}\sigma_{l}) & \Leftrightarrow & -\delta_{jk}+\varepsilon_{jkl}e_{l})\,\,(\forall\,\,\, j,k,l=1,2,3).\label{eq:25}\end{eqnarray}
As such,in contrast to $\mathbb{H}$, the Cayley algebra $\mathcal{O}$
cannot be represented by matrices with the usual multiplication rules
due to its non associative nature. However, it is possible to represent
octonions by matrices, provided one defines a special multiplication
rule among them in terms of its split octonion basis elements.

\section{Split Octonions}

The split octonions are a non associative extension of quaternions
(or the split quaternions).They differ from the octonion in the signature
of quadratic form. the split octonions have a signature $(4,4)$ whereas
the octonions have positive signature $(8,0)$. The Cayley algebra
of octonions over the field of complex numbers $\mathbb{C_{C}=C\otimes}C$
is visualized as the algebra of split octonions with its following
basis elements,

\begin{eqnarray}
u_{0}=\frac{1}{2} & (1+i\,\, e_{7}),\,\,\,\,\,\,\,\,\,\,\, & u_{0}^{\star}=\frac{1}{2}(1-i\,\, e_{7}),\nonumber \\
u_{1}=\frac{1}{2} & (e_{1}+i\,\, e_{4}),\,\,\,\,\,\,\,\,\,\,\, & u_{1}^{\star}=\frac{1}{2}(e_{1}-i\,\, e_{4}),\nonumber \\
u_{2}=\frac{1}{2} & (e_{2}+i\,\, e_{5}),\,\,\,\,\,\,\,\,\,\,\, & u_{2}^{\star}=\frac{1}{2}(e_{2}-i\,\, e_{5}),\nonumber \\
u_{3}=\frac{1}{2} & (e_{3}+i\,\, e_{6}),\,\,\,\,\,\,\,\,\,\,\, & u_{3}^{\star}=\frac{1}{2}(e_{3}-i\,\, e_{6}),\label{eq:26}\end{eqnarray}
where $(i=\sqrt{-1})$ is usual complex imaginary number and commutes
with all the seven octonion imaginary units $e_{A}(A=1,2...,7)$.
Using the multiplication table of octonion we get the following multiplication
table for split octonion basis elements $u_{\beta}$and $u_{\beta}^{\star}(\beta=0,1,2,3)$
as,

\begin{tabular}{|c|c|c|c|c|c|c|c|c|}
\hline 
$.$&
$u_{0}^{\star}$&
$u_{1}^{\star}$&
$u_{2}^{\star}$&
$u_{3}^{\star}$&
$u_{0}$&
$u_{1}$&
$u_{2}$&
$u_{3}$\tabularnewline
\hline
\hline 
$u_{0}^{\star}$&
$u_{0}^{\star}$&
$u_{1}^{\star}$&
$u_{2}^{\star}$&
$u_{3}^{\star}$&
$0$&
$0$&
$0$&
$0$\tabularnewline
\hline 
$u_{1}^{\star}$&
$0$&
$0$&
$u_{3}$&
$-u_{2}$&
$u_{1}^{\star}$&
$-u_{0}^{\star}$&
$0$&
$0$\tabularnewline
\hline 
$u_{2}^{\star}$&
$0$&
$-u_{3}$&
$0$&
$u_{1}$&
$u_{2}^{\star}$&
$0$&
$-u_{0}^{\star}$&
$0$\tabularnewline
\hline 
$u_{3}^{\star}$&
$0$&
$u_{2}$&
$-u_{1}$&
$0$&
$u_{3}^{\star}$&
$0$&
$0$&
$-u_{0}^{\star}$\tabularnewline
\hline 
$u_{0}$&
$0$&
$0$&
$0$&
$0$&
$u_{0}$&
$u_{1}$&
$u_{2}$&
$u_{3}$\tabularnewline
\hline 
$u_{1}$&
$u_{1}$&
$-u_{0}$&
$0$&
$0$&
$0$&
$0$&
$u_{3}^{\star}$&
$-u_{2}^{\star}$\tabularnewline
\hline 
$u_{2}$&
$u_{2}$&
$0$&
$-u_{0}$&
$0$&
$0$&
$-u_{3}^{\star}$&
$0$&
$u_{1}^{\star}$\tabularnewline
\hline 
$u_{3}$&
$u_{3}$&
$0$&
$0$&
$-u_{0}$&
$0$&
$u_{2}^{\star}$&
$-u_{1}^{\star}$&
$0$\tabularnewline
\hline
\end{tabular}

\begin{description}
\item [{~~~~~~~~~~~Table2-}] Split Octonion Multiplication table
\end{description}
As such an octonion $Z$ can be expressed in terms of split octonion
basis elements as 

\begin{eqnarray}
Z & =z_{o}u_{0}+z_{o}^{\star}u_{0}^{\star} & +z_{j}u_{j}+z_{j}^{\star}u_{j}^{\star}\label{eq:27}\end{eqnarray}
where $z_{\beta}(\beta=0,1,2,3)$ are the complex numbers and 

\begin{eqnarray}
u_{0}Z=z_{\beta}u_{\beta}=z & ;\,\,\,\,\,\,\,\,\,\,\, & u_{0}^{\star}Z=z_{\beta}^{\star}u_{\beta}^{\star}=z^{\star}.\label{eq:28}\end{eqnarray}
Like octonions , split octonions are non commutative and non associative
. They also form a composition algebra and satisfy equation (\ref{eq:18}).
Split octonions also satisfy the Moufang identities and thus form
the alternative algebra. Therefore,by Artin's theorem, the sub algebra
generated by any two elements is associative and the set of all invertible
elements (i.e. those elements for which $N(x)\neq0)$ describe a Moufang
loop. We may now introduce a convenient realization for the split
octonion basis elements $(u_{0},u_{0}^{\star},u_{j},u_{j}^{\star})(j=1,2,3)$
in terms of quaternion basis elements, $e_{0}\rightarrow\sigma_{0}=\hat{1}$
and $e_{j}\rightarrow-i\sigma_{j}$as 

\begin{eqnarray}
u_{0}=\left(\begin{array}{cc}
0 & 0\\
0 & 1\end{array}\right) & ; & \,\,\,\,\,\,\,\,\,\,\,\,\,\,\, u_{0}^{\star}=\left(\begin{array}{cc}
1 & 0\\
0 & 0\end{array}\right)\nonumber \\
u_{j}=\left(\begin{array}{cc}
0 & 0\\
e_{j} & 0\end{array}\right) & ; & \,\,\,\,\,\,\,\,\,\,\,\,\,\,\, u_{j}^{\star}=\left(\begin{array}{cc}
0 & -e_{j}\\
0 & 0\end{array}\right).\label{eq:29}\end{eqnarray}
The split Cayley (octonion) algebra is thus expressed in terms of
$2\times2$ Zorn's vector matrices components of which are scalar
and vector parts of a quaternion i.e.

\begin{eqnarray}
\mathcal{O} & =\{ & \left(\begin{array}{cc}
m & \overrightarrow{p}\\
\overrightarrow{q} & n\end{array}\right):\,\,\, m,n\in Sc(\mathbb{H});\,\,\,\overrightarrow{p},\overrightarrow{q}\in Vec(\mathbb{H})\}.\label{eq:30}\end{eqnarray}
As such , we may also write an arbitrary split octonion $A$ in terms
of following $2\times2$ Zorn's vector matrix realizations as.

\begin{eqnarray}
A=au_{0}^{\star}+bu_{0}+x_{j}u_{j}^{\star}+y_{j}u_{j} & = & \left(\begin{array}{cc}
a & -\overrightarrow{x}\\
\overrightarrow{y} & b\end{array}\right)\label{eq:31}\end{eqnarray}
where $a$ and $b$ are scalars and $\overrightarrow{x}$ and $\overrightarrow{y}$
are three vectors. Thus the product of two octonions in terms of $2\times2$
Zorn's vector matrix realization is expressed as 

\begin{eqnarray}
\left(\begin{array}{cc}
a & \overrightarrow{x}\\
\overrightarrow{y} & b\end{array}\right)\left(\begin{array}{cc}
c & \overrightarrow{u}\\
\overrightarrow{v} & d\end{array}\right) & = & \left(\begin{array}{cc}
ac+\overrightarrow{x}\cdot\overrightarrow{v} & a\overrightarrow{u}+d\overrightarrow{x}-\overrightarrow{y}\times\overrightarrow{v}\\
c\overrightarrow{y}+b\overrightarrow{v}+\overrightarrow{x}\times\overrightarrow{u} & \overrightarrow{y}\cdot\overrightarrow{u}+bd\end{array}\right)\label{eq:32}\end{eqnarray}
where $(\times)$ denotes the usual vector product, $e_{j}$ $(j=1,2,3)$with
$e_{j}\times e_{k}=\varepsilon_{jkl}e_{l}$and $e_{j}e_{k}=-\delta_{jk}$.
As such, we can relate the split octonions to the vector matrices
given by equation (\ref{eq:29}). Octonion conjugate of equation(\ref{eq:31})
in terms of $2\times2$ Zorn's vector matrix realization is now defined
as

\begin{eqnarray}
\overline{A}=bu_{0}^{\star}+au_{0}-x_{j}u_{j}^{\star}-y_{j}u_{j} & = & \left(\begin{array}{cc}
b & \overrightarrow{x}\\
-\overrightarrow{y} & a\end{array}\right).\label{eq:33}\end{eqnarray}
The norm of $A$is then defined as,

\begin{eqnarray}
N(A)=A\,\overline{A}=\overline{A}\, A & = & (ab+\overrightarrow{x}\,\cdot\overrightarrow{y})\cdot\hat{1}=n(A)\hat{1}\label{eq:34}\end{eqnarray}
where $\hat{1}$ is the identity element of the algebra given by $\hat{1}=1u_{0}^{\star}+1u_{0}$
and the expression $n(A)=(ab+\overrightarrow{x}\,\cdot\overrightarrow{y})$
defines the quadratic form which admits the composition $n(\overrightarrow{A}.\overrightarrow{B})=n(\overrightarrow{A})\, n(\overrightarrow{B})$
for all $\overrightarrow{A}$,$\overrightarrow{B}\in\mathcal{O}$.
As such, we may easily express the Euclidean or Minikowski four vector
in split octonion formulation in terms of $2\times2$ Zorn's vector
matrix realization. So, the space-time four-differential operator
and its conjugates are then be written as

\begin{eqnarray*}
\boxdot & = & \left(\begin{array}{cc}
\partial_{4} & -\overrightarrow{\nabla}\\
\overrightarrow{\nabla} & \partial_{4}\end{array}\right)=\left(\begin{array}{cc}
i\partial_{0} & -\overrightarrow{\nabla}\\
\overrightarrow{\nabla} & i\partial_{0}\end{array}\right);\end{eqnarray*}

\begin{eqnarray}
\overline{\boxdot} & = & \left(\begin{array}{cc}
\partial_{4} & \overrightarrow{\nabla}\\
-\overrightarrow{\nabla} & \partial_{4}\end{array}\right)=\left(\begin{array}{cc}
i\partial_{0} & \overrightarrow{\nabla}\\
-\overrightarrow{\nabla} & i\partial_{0}\end{array}\right).\label{eq:35}\end{eqnarray}

\section{Duality Invariance and Generalized Fields of Dyons and Gravito-dyons}

~~~~~~~Duality invariance is an old idea introduced a century
ago in classical electromagnetism for Maxwell's equations in vacuum
i.e.

\begin{eqnarray}
\overrightarrow{\nabla}\cdot\overrightarrow{E}=0\,\,\,\,\,\,\,\,\,\,\,\,\,\, & ;\,\,\,\,\, & \overrightarrow{\nabla}\times\overrightarrow{E}=-\frac{\partial\overrightarrow{M}}{\partial t}\nonumber \\
\overrightarrow{\nabla}\cdot\overrightarrow{M}=0\,\,\,\,\,\,\,\,\,\,\,\,\,\, & ;\,\,\,\,\, & \overrightarrow{\nabla}\times\overrightarrow{M}=\frac{\partial\overrightarrow{E}}{\partial t}.\label{eq:36}\end{eqnarray}
where $\overrightarrow{E}$ is the electric field and $\overrightarrow{M}$
is the magnetic fields. For brevity we have made use of the natural
units $(c=\hbar=1)$, and taking the other constants like gravitational
constant as unity though out the text. Maxwell's equations in vacuum
are symmetrical as well as invariant under both Lorentz transformations
(in fact, conformal) and electromagnetic duality transformations given
by,

\begin{eqnarray}
\overrightarrow{E}\,\rightarrow\overrightarrow{E}\,\cos\theta+\overrightarrow{M}\,\sin\theta;\,\,\,\,\,\,\,\,\, & \overrightarrow{M} & \,\rightarrow-\overrightarrow{E}\,\sin\theta+\overrightarrow{M}\,\cos\theta.\label{eq:37}\end{eqnarray}
 For a particular value of $\theta=\frac{\pi}{2}$, equation (\ref{eq:37})
reduces to,

\begin{eqnarray}
\overrightarrow{E}\,\rightarrow\overrightarrow{M};\,\,\,\, & \overrightarrow{M} & \,\rightarrow-\overrightarrow{E}\,\,\,\,\,\,\,\,\, or\nonumber \\
\left(\begin{array}{c}
\overrightarrow{E}\\
\overrightarrow{M}\end{array}\right) & = & \left(\begin{array}{cc}
0 & -1\\
1 & 0\end{array}\right)\left(\begin{array}{c}
\overrightarrow{E}\\
\overrightarrow{M}\end{array}\right).\label{eq:38}\end{eqnarray}
In terms of complex vectors the duality transformations are visualized
as 

\begin{eqnarray}
(\overrightarrow{E}+i\overrightarrow{M})\rightarrow & \exp(i\theta) & (\overrightarrow{E}+i\overrightarrow{M})\label{eq:39}\end{eqnarray}
Lorentz invariance is obeyed even if we write Maxwell's equations
in covariant formulation on introducing the electromagnetic field
strengths. The duality symmetry is lost if electric charge and current
source densities enter to Maxwell's equations. However, conventional
Maxwell's equations are invariant under Lorentz and conformal transformations
but neither these are symmetrical nor are invariant under the duality
transformations (\ref{eq:37}, \ref{eq:38}, \ref{eq:39}) . Dirac
\cite{key-1} put forward this idea and introduced the concept of
magnetic monopole not only to symmetrize the Maxwell's equations but
also to make them dual invariant. Thus, electromagnetic duality requires:-

\begin{itemize}
\item The existence of magnetic monopoles
\item The existence of magnetic monopole is closely related to the existence
of a compact  $U(1)$gauge group.
\item The magnetic charge implies the C- invariance 
\item Monopole equations are to be invariant under duality transformations.
\end{itemize}
Consequently, the Generalized Dirac Maxwell's (GDM )equations given
below

\begin{eqnarray}
\overrightarrow{\nabla}\cdot\overrightarrow{E}=\rho_{e}; & \overrightarrow{\nabla}\times\overrightarrow{E}= & -\frac{\partial\overrightarrow{M}}{\partial t}-\overrightarrow{j_{m}};\nonumber \\
\overrightarrow{\nabla}\cdot\overrightarrow{M}=\rho_{m}; & \nabla\times\overrightarrow{M}= & \overrightarrow{j_{e}}+\frac{\partial\overrightarrow{E}}{\partial t};\label{eq:40}\end{eqnarray}
are invariant under duality transformations (\ref{eq:37}, \ref{eq:38},
\ref{eq:39}) incorporating the following duality among the electric
and magnetic charge and current source densities 

\begin{eqnarray*}
\left(\begin{array}{c}
\rho_{e}\\
\rho_{m}\end{array}\right) & = & \left(\begin{array}{cc}
0 & -1\\
1 & 0\end{array}\right)\left(\begin{array}{c}
\rho_{e}\\
\rho_{m}\end{array}\right);\end{eqnarray*}

\begin{eqnarray}
\left(\begin{array}{c}
\overrightarrow{j_{e}}\\
\overrightarrow{J_{m}}\end{array}\right) & = & \left(\begin{array}{cc}
0 & -1\\
1 & 0\end{array}\right)\left(\begin{array}{c}
\overrightarrow{j_{e}}\\
\overrightarrow{J_{m}}\end{array}\right).\label{eq:41}\end{eqnarray}
 Here $\rho_{e}$ is the electric charge source density, $\rho_{m}$
is magnetic charge (monopole) source density, $\overrightarrow{j_{e}}$is
the electric current source density and $\overrightarrow{j_{m}}$
is magnetic current (monopole) source density. 

Accordingly, on postulating the existence of Heavisidian monopole
\cite{key-20,key-21,key-22,key-23} and keeping in view the asymmetry
therein between the gravitational (gravi-electric) and Heavisidian
(gravi-magnetic) in Maxwellian gravity, the structural symmetry between
these two interactions describes the invariance of GDM type equations
for gravito-Heavisidian fields 

\begin{eqnarray}
\overrightarrow{\nabla}\cdot\overrightarrow{G}=\rho_{g}; & \overrightarrow{\nabla}\times\overrightarrow{G}= & -\frac{\partial\overrightarrow{H}}{\partial t}-\overrightarrow{j_{h}};\nonumber \\
\overrightarrow{\nabla}\cdot\overrightarrow{H}=\rho_{m}; & \nabla\times\overrightarrow{H}= & \overrightarrow{j_{h}}+\frac{\partial\overrightarrow{G}}{\partial t};\label{eq:42}\end{eqnarray}
under the following duality transformations,

\begin{eqnarray*}
\left(\begin{array}{c}
\overrightarrow{G}\\
\overrightarrow{H}\end{array}\right) & = & \left(\begin{array}{cc}
0 & -1\\
1 & 0\end{array}\right)\left(\begin{array}{c}
\overrightarrow{G}\\
\overrightarrow{H}\end{array}\right);\end{eqnarray*}

\begin{eqnarray}
\left(\begin{array}{c}
\rho_{g}\\
\rho_{h}\end{array}\right) & = & \left(\begin{array}{cc}
0 & -1\\
1 & 0\end{array}\right)\left(\begin{array}{c}
\rho_{g}\\
\rho_{h}\end{array}\right);\nonumber \\
\left(\begin{array}{c}
\overrightarrow{j_{g}}\\
\overrightarrow{J_{h}}\end{array}\right) & = & \left(\begin{array}{cc}
0 & -1\\
1 & 0\end{array}\right)\left(\begin{array}{c}
\overrightarrow{j_{g}}\\
\overrightarrow{J_{h}}\end{array}\right).\label{eq:43}\end{eqnarray}
here $\overrightarrow{G}$ is the gravitational (gravi-electric) field
, $\overrightarrow{H}$ is the Heavisidian (gravi-magnetic) fields,
$\rho_{g}$ is the gravitational (gravi-electric) charge (mass) density,
$\rho_{m}$ is Heavisidian (gravi-magnetic) monopole (mass) density,
$\overrightarrow{j_{g}}$is the gravitational (gravi-electric) current
density and $\overrightarrow{j_{h}}$ is Heavisidian (gravi-magnetic)
monopole current density in Maxwellian gravity (namely linear gravity).
We may write these theories in covariant formulations accordingly
by introducing the corresponding field strengths and gauge potentials.
But the introduction of magnetic (Heavisidian) monopole (mass) leads
to various discrepancies like the string singularity and even the
Dirac quantization condition is no more dual invariant. Then the theories
of dyons (particles carrying simultaneous existence of electric and
magnetic charges) come in force and Dirac quantization condition have
been made dual invariant by replacing it with Schwinger-Zwanziger
\cite{key-2,key-3} quantization condition. In order to avoid the
use of arbitrary string variables and keeping in mind the idea of
two four-potentials, we have developed a manifestly covariant and
dual invariant theory of generalized electromagnetic fields of dyons
\cite{key-12,key-13} and accordingly those of generalized gravito-Heavisidian
fields of gravito-dyons \cite{key-23,key-24} on assuming the generalized
charge, four-potential,vector field, current and generalized field
tensors of dyons ( gravito dyons ) as a complex (order pair of two
real numbers) ones like equation (\ref{eq:39}) with their real and
imaginary parts as a electric (gravitational) and magnetic (Heavisidian)constituents.
Hence the Generalized Dirac Maxwell's (GDM )equations given by equations
(\ref{eq:40}) and (\ref{eq:42}) are respectively the field equations
of dyons and gravito-dyons. Let us summaries these two theories in
the following table;

\begin{flushleft}
\begin{tabular}{|l|l|l|}
\hline 
Dynamical variables &
Fields associated with dyons&
Fields associated with Gravito-dyons\tabularnewline
\hline
\hline 
Generalized Charge (mass)&
$Q^{EM}=(e,g)=(e+ig)$&
$Q^{GH}=(m,\, h)=(m+ih)$\tabularnewline
\hline 
Generalized four-potential&
$V_{\mu}^{EM}=(A_{\mu},B_{\mu})=(A_{\mu}+iB_{\mu})$&
$V_{\mu}^{GH}=(C_{\mu},D_{\mu})=(C_{\mu}+iD_{\mu})$\tabularnewline
\hline 
Generalized four-current &
$J_{\mu}^{EM}=(j_{\mu}^{(E)},\, j_{\mu}^{(M)})=(j_{\mu}^{(E)}+i\, j_{\mu}^{(M)})$&
$J_{\mu}^{GH}=(\, j_{\mu}^{(G)},j_{\mu}^{(H)}\,)$\tabularnewline
\hline 
Generalized vector- field&
$\overrightarrow{\psi}^{EM}=(\overrightarrow{E},\overrightarrow{M})=(\overrightarrow{E},+i\,\overrightarrow{M})$&
$\overrightarrow{\psi}^{GH}=(\overrightarrow{G},\overrightarrow{H})=(\overrightarrow{G}+i\,\overrightarrow{H})$\tabularnewline
\hline 
Generalized field Tensor&
$F_{\mu\nu}^{EM}=(A_{\mu\nu},\, B_{\mu\nu})=(A_{\mu\nu}+i\, B_{\mu\nu})$&
$F_{\mu\nu}^{GH}=(C_{\mu\nu},\, D_{\mu\nu})=(C_{\mu\nu}+i\, D_{\mu\nu})$\tabularnewline
\hline
\end{tabular}
\par\end{flushleft}

where $EM$ stands for electromagnetic, $GH$ is used for gravito-Heavisidian;
$A_{\mu\nu}=A_{\mu,\nu}-A_{\nu,\mu}=\partial_{\nu}A_{\mu}-\partial_{\mu}A_{\nu}$;
$B_{\mu\nu}=B_{\mu,\nu}-B_{\nu,\mu}=\partial_{\nu}B_{\mu}-\partial_{\mu}B_{\nu}$
; $C_{\mu\nu}=C_{\mu,\nu}-C_{\nu,\mu}=\partial_{\nu}C_{\mu}-\partial_{\mu}C_{\nu}$and
$D_{\mu\nu}=D_{\mu,\nu}-D_{\nu,\mu}=\partial_{\nu}D_{\mu}-\partial_{\mu}D_{\nu}(\mu,\nu=0,1,2,3).$
Here the real parameters of complex variables are described as the
electric (gravitational) constituents while the imaginary counter
parts of complex variables are identified as the magnetic (Heavisidian)
constituents of dyons (gravito-dyons). These fields and their quantum
equations will described in detail in covariant formulation in the
next sections with the applications of split octonions. As such, duality
transformations for these dynamical variables associated with dyons
(gravito-dyons) in generalized electromagnetic (gravito-Heavisidian)
fields take the following forms,

\begin{eqnarray}
Q & \rightarrow & (\exp\, i\theta).(Q)\nonumber \\
V_{\mu} & \rightarrow & (\exp\, i\theta).(V_{\mu})\nonumber \\
J_{\mu} & \rightarrow & (\exp\, i\theta).(J_{\mu})\nonumber \\
\overrightarrow{\psi} & \rightarrow & (\exp\, i\theta).(\overrightarrow{\psi})\nonumber \\
F_{\mu\nu} & \rightarrow & (\exp\, i\theta).(F_{\mu\nu}).\label{eq:44}\end{eqnarray}
Hence with these transformations the GDM equations (\ref{eq:40},
\ref{eq:42}), corresponding covariant field equations , equation
of motion, Schwinger-Zwanziger \cite{key-2,key-3} quantization condition,
BPS mass formula and the energy- momentum densities of generalized
electromagnetic ((gravito-Heavisidian)) fields of dyons (gravito-dyons)
in complex representation are invariant. The duality conjecture has
now been gaining enormous potential importance in connection with
latest developments of elementary particles in gauge theories, grand
unified theories, supersymmetry and super strings.

\section{Split Octonion Formulation for Unified Fields of Dyons}

~~~~~~~We may apply now the split octonion formalism in order
to formulate the unified theory of generalized electromagnetic fields
of dyons \cite{key-12,key-13} and those of generalized gravito-Heavisidian
fields of gravito-dyons \cite{key-23,key-24} discussed above. Let
us combine the generalized charges of dyons and gravito-dyons (i.e.
both complex quantities) with the help of Cayley Dickson process to
make them an unified quaternion tetrad. As such we may express the
quaternion charge for the unified fields of dyons and gravito-dyons
as 

\begin{eqnarray}
Q= & (e+i\, g)+(m+i\, h)j=e+i\, g+j\, m+i\, j\, h=e+i\, g+j\, m+k\, h= & (e\,,g\,,m\,,h)\label{eq:45}\end{eqnarray}
where $i$,$\, j$, $k$ are the three non commutating quaternion
imaginary elements $i^{2}=J^{2}=k^{2}=-1;$$\,\,\, i\, j=-j\, i=k;\,\,\, j\, k=-k\, j=i;\,\,\, k\, i=-i\, k=j$
and we may replace them by the quaternion units $e_{1}$,$\, e_{2}$
and $e_{3}$ of quaternion tetrad $(1,\, e_{1},\, e_{2},\, e_{3})$
which satisfies the multiplication rules given by equation (\ref{eq:25}).
Unfortunately the quaternion basis elements loose the matrix realization
when we write them in split basis. Like equation (\ref{eq:31}), we
may now define the split octonion representation of unified quaternion
charge of dyons and gravito-dyons in terms of $2\times2$ Zorn's vector
matrix realization as

\begin{eqnarray*}
Q= & (e,g,m,h) & =\left\{ \begin{array}{cc}
e & -e_{1}g-e_{2}m-e_{3}h\\
e_{1}g+e_{2}m+e_{3}h & e\end{array}\right\} \end{eqnarray*}

\begin{eqnarray}
= & e(u_{0}^{\star}+u_{0})+g(u_{1}+u_{1}^{\star})+ & m(u_{2}+u_{2}^{\star})+h(u_{3}+u_{3}^{\star}).\label{eq:46}\end{eqnarray}
Like equation (\ref{eq:33}), we may write split octonion conjugate
of unified quaternion charge of dyons and gravito-dyons in terms of
$2\times2$ Zorn's vector matrix realization as 

\begin{eqnarray}
\overline{Q}= & (e,-g,-m,-h) & =\left\{ \begin{array}{cc}
e & e_{1}g+e_{2}m+e_{3}h\\
-e_{1}g-e_{2}m-e_{3}h & e\end{array}\right\} \nonumber \\
= & e(u_{0}^{\star}+u_{0})-g(u_{1}+u_{1}^{\star})- & m(u_{2}+u_{2}^{\star})-h(u_{3}+u_{3}^{\star}).\label{eq:47}\end{eqnarray}
The norm of split octonion form of unified quaternion charge of dyons
and gravito-dyons is defined as,

\begin{eqnarray}
N(Q) & =Q\overline{Q} & =\overline{Q}Q\nonumber \\
=\left(\begin{array}{cc}
e^{2}+g^{2}+m^{2}+h^{2} & 0\\
0 & e^{2}+g^{2}+m^{2}+h^{2}\end{array}\right).\hat{1} & = & (e^{2}+g^{2}+m^{2}+h^{2}).\hat{1}.\label{eq:48}\end{eqnarray}
The interaction of $a^{th}$ split octonionic charge $Q_{a}$ in the
field of other $b^{th}$split octonionic charge $Q_{b}$ now depends
on the quantity,

\begin{eqnarray}
Q_{a}.\overline{Q_{b}} & = & u_{0}(e_{a}e_{b}+m_{a}m_{b}+g_{a}g_{b}+h_{a}h_{b})\nonumber \\
+ & u_{1} & (-e_{a}g_{b}+g_{a}e_{b}+m_{a}h_{b}-h_{a}m_{b})\nonumber \\
+ & u_{2} & (-e_{a}m_{b}+m_{a}e_{b}+h_{a}g_{b}-g_{a}h_{b})\nonumber \\
+ & u_{3} & (-e_{a}h_{b}+h_{a}e_{b}+g_{a}m_{b}-m_{a}g_{b})\nonumber \\
+ & u_{0}^{\star} & (e_{a}e_{b}+m_{a}m_{b}+g_{a}g_{b}+h_{a}h_{b})\nonumber \\
+ & u_{1}^{\star} & (h_{a}m_{b}-m_{a}h_{b}-e_{a}g_{b}+g_{a}e_{b})\nonumber \\
+ & u_{2}^{\star} & (g_{a}h_{b}-h_{a}g_{b}-e_{a}m_{b}+m_{a}e_{b})\nonumber \\
+ & u_{3}^{\star} & (m_{a}g_{b}-g_{a}m_{b}-e_{a}h_{b}+h_{a}e_{b})\label{eq:49}\end{eqnarray}
and 

\begin{eqnarray}
\overline{Q_{a}}.Q_{b} & = & u_{0}(e_{a}e_{b}+m_{a}m_{b}+g_{a}g_{b}+h_{a}h_{b})\nonumber \\
+ & u_{1} & (e_{a}g_{b}-g_{a}e_{b}+m_{a}h_{b}-h_{a}m_{b})\nonumber \\
+ & u_{2} & (e_{a}m_{b}-m_{a}e_{b}+h_{a}g_{b}-g_{a}h_{b})\nonumber \\
+ & u_{3} & (e_{a}h_{b}-h_{a}e_{b}+g_{a}m_{b}-m_{a}g_{b})\nonumber \\
+ & u_{0}^{\star} & (e_{a}e_{b}+m_{a}m_{b}+g_{a}g_{b}+h_{a}h_{b})\nonumber \\
+ & u_{1}^{\star} & (h_{a}m_{b}-m_{a}h_{b}+e_{a}g_{b}-g_{a}e_{b})\nonumber \\
+ & u_{2}^{\star} & (g_{a}h_{b}-h_{a}g_{b}+e_{a}m_{b}-m_{a}e_{b})\nonumber \\
+ & u_{3}^{\star} & (m_{a}g_{b}-g_{a}m_{b}+e_{a}h_{b}-h_{a}e_{b}).\label{eq:50}\end{eqnarray}
Like equation (\ref{eq:31}), we may now introduce the unified split
octonion form of quaternion valued four-potential of dyons and gravito-dyons
in terms of $2\times2$ Zorn's vector matrix realization as,

\begin{flushleft}
\begin{eqnarray}
V & = & \left(\begin{array}{cc}
A_{0}+B_{0}+C_{0}+D_{0} & -(\overrightarrow{A}+\overrightarrow{B}+\overrightarrow{C}+\overrightarrow{D})\\
(\overrightarrow{A}+\overrightarrow{B}+\overrightarrow{C}+\overrightarrow{D}) & A_{0}+B_{0}+C_{0}+D_{0}\end{array}\right)=\left(\begin{array}{cc}
V_{0} & -\overrightarrow{V}\\
\overrightarrow{V} & V_{0}\end{array}\right)\label{eq:51}\end{eqnarray}
where $A$, $B$,$C$ and $D$ are the quaternionic forms of four-potential
associated with electric, magnetic, gravitational (g-electric) and
Heavisidian (g-magnetic) charges respectively. These are also written
as follows in split octonionic formulation in terms of $2\times2$
Zorn's vector matrix realization ,
\par\end{flushleft}

\begin{eqnarray}
A & =\left(\begin{array}{cc}
A_{0} & -\overrightarrow{A}\\
\overrightarrow{A} & A_{0}\end{array}\right)= & \left(\begin{array}{cc}
A_{0} & -(A_{1}e_{1}+A_{2}e_{2}+A_{3}e_{3})\\
(A_{1}e_{1}+A_{2}e_{2}+A_{3}e_{3}) & A_{0}\end{array}\right);\nonumber \\
B & =\left(\begin{array}{cc}
B_{0} & -\overrightarrow{B}\\
\overrightarrow{B} & B_{0}\end{array}\right)= & \left(\begin{array}{cc}
B_{0} & -(B_{1}e_{1}+B_{2}e_{2}+B_{3}e_{3})\\
(B_{1}e_{1}+B_{2}e_{2}+B_{3}e_{3}) & B_{0}\end{array}\right);\nonumber \\
C & =\left(\begin{array}{cc}
C_{0} & -\overrightarrow{C}\\
\overrightarrow{C} & C_{0}\end{array}\right)= & \left(\begin{array}{cc}
C_{0} & -(C_{1}e_{1}+C_{2}e_{2}+C_{3}e_{3})\\
(C_{1}e_{1}+C_{2}e_{2}+C_{3}e_{3}) & C_{0}\end{array}\right);\nonumber \\
D & =\left(\begin{array}{cc}
D_{0} & -\overrightarrow{D}\\
\overrightarrow{D} & D_{0}\end{array}\right)= & \left(\begin{array}{cc}
D_{0} & -(D_{1}e_{1}+D_{2}e_{2}+D_{3}e_{3})\\
(D_{1}e_{1}+D_{2}e_{2}+D_{3}e_{3}) & D_{0}\end{array}\right).\label{eq:52}\end{eqnarray}
As such, we have reformulated the four vector potentials of all the
individual charges namely, electric, magnetic, gravitational (g-electric)
and Heavisidian (g-magnetic), by the virtue of split octonion analyticity
in terms of $2\times2$ Zorn's vector matrix realization. Accordingly,
the split octonion form of quaternion unified vector field of dyons
and gravito-dyons may then be expressed as

\begin{eqnarray}
\overrightarrow{\Psi} & = & \left(\begin{array}{cc}
0 & -(\overrightarrow{E}+\overrightarrow{M}+\overrightarrow{G}+\overrightarrow{H})\\
(\overrightarrow{E}+\overrightarrow{M}+\overrightarrow{G}+\overrightarrow{H}) & 0\end{array}\right)\label{eq:53}\end{eqnarray}
where $C$ , $\overrightarrow{M}$, $\overrightarrow{G}$ and $\overrightarrow{H}$
are respectively the generalized electric, magnetic, gravitational
and Heavisidian fields described in terms of two four potential theory
of dyons and gravito-dyons with the following definitions ,

\begin{eqnarray}
\overrightarrow{E} & =- & \frac{\partial\overrightarrow{A}}{\partial t}-\overrightarrow{\nabla}A_{0}-\overrightarrow{\nabla}\times\overrightarrow{B};\nonumber \\
\overrightarrow{M} & = & -\frac{\partial\overrightarrow{B}}{\partial t}-\overrightarrow{\nabla}B_{0}+\overrightarrow{\nabla}\times\overrightarrow{A};\nonumber \\
\overrightarrow{G} & = & -\frac{\partial\overrightarrow{C}}{\partial t}-\overrightarrow{\nabla}C_{0}+\overrightarrow{\nabla}\times\overrightarrow{D};\nonumber \\
\overrightarrow{H} & = & \frac{\partial\overrightarrow{D}}{\partial t}+\overrightarrow{\nabla}D_{0}+\overrightarrow{\nabla}\times\overrightarrow{C};\label{eq:54}\end{eqnarray}
where $\overrightarrow{\nabla}=(\partial_{1},\partial_{2},\partial_{3})=(e_{1}\partial_{1}+e_{2}\partial_{2},+e\partial_{3})$
and these generalized electric, magnetic, gravitational and Heavisidian
fields satisfy the pairs of Generalized Dirac-Maxwell's (GDM) equations
(\ref{eq:40}, \ref{eq:42}) of dyons and gravito-dyons. the components
of split octonion valued unified four potential $V$ given by equation
(\ref{eq:51}) and unified field $\overrightarrow{\Psi}$ given by
equation (\ref{eq:53}) establish the following relation among them
as ,

\begin{eqnarray}
\overrightarrow{\Psi} & = & -\frac{\partial\overrightarrow{V}}{\partial t}-\overrightarrow{\nabla}V_{0}+i\overrightarrow{\nabla}\times\overrightarrow{V}.\label{eq:55}\end{eqnarray}
Operating $\boxdot$ given by equation (\ref{eq:35}) to equation
(\ref{eq:51}) and using the properties of multiplication of split
octonion algebra n terms of $2\times2$ Zorn's vector matrix realization,
we get the following split octonion form of potential field equation
for the unified theory of dyons and gravito-dyons as,

\begin{eqnarray}
\boxdot V & = & \Psi\label{eq:56}\end{eqnarray}
where

\begin{eqnarray*}
\Psi & = & \left(\begin{array}{cc}
\partial_{0}V_{0}+\overrightarrow{\nabla}\cdot\overrightarrow{V} & \frac{\partial\overrightarrow{V}}{\partial t}+\overrightarrow{\nabla}V_{0}-i\overrightarrow{\nabla}\times\overrightarrow{V}\\
-\frac{\partial\overrightarrow{V}}{\partial t}-\overrightarrow{\nabla}V_{0}+i\overrightarrow{\nabla}\times\overrightarrow{V} & \partial_{0}V_{0}+\overrightarrow{\nabla}\cdot\overrightarrow{V}\end{array}\right)=\left(\begin{array}{cc}
0 & -\overrightarrow{\Psi}\\
\overrightarrow{\Psi} & 0\end{array}\right)\end{eqnarray*}
is the split octonion form of unified vector field while the diagonal
components vanish due to the Lorentz gauge conditions applied to each
four potentials. Equation (\ref{eq:56}) is the split octonion potential
wave equation for the unified fields of dyons and those of gravito
dyons.This equation may also be visualized as the analogue of unified
GDM equations of dyons and gravito-dyons and is invariant under duality,
quaternion and Lorentz transformations. As such, the unified potential
field equation (\ref{eq:56}) is simple, compact ,consistent and manifestly
covariant. Accordingly, we may write the split octonion representation
for four-current associated with the unified fields of dyons and gravito-dyons
, the components of which are given by Eq. (\ref{eq:40}\ref{eq:42}),
in the following manner ,

\begin{eqnarray}
J & = & \left(\begin{array}{cc}
\rho_{e}+\rho_{g}+\rho_{m}+\rho_{h} & -(\overrightarrow{j_{e}}+\overrightarrow{j_{g}}+\overrightarrow{j_{m}}+\overrightarrow{j_{h}})\\
(\overrightarrow{j_{e}}+\overrightarrow{j_{g}}+\overrightarrow{j_{m}}+\overrightarrow{j_{h}}) & \rho_{e}+\rho_{g}+\rho_{m}+\rho_{h}\end{array}\right)=\left(\begin{array}{cc}
J_{0} & -\overrightarrow{J}\\
\overrightarrow{J} & J_{0}\end{array}\right)\label{eq:57}\end{eqnarray}
where

\begin{eqnarray}
\boxdot\overline{\boxdot}A=\overline{\boxdot}\,\boxdot A= & j_{e} & =\left(\begin{array}{cc}
\rho_{e} & -\overrightarrow{j_{e}}\\
\overrightarrow{j_{e}} & \rho_{e}\end{array}\right);\nonumber \\
\boxdot\overline{\boxdot}B=\overline{\boxdot}\,\boxdot B= & j_{g} & =\left(\begin{array}{cc}
\rho_{g} & -\overrightarrow{j_{g}}\\
\overrightarrow{j_{g}} & \rho_{g}\end{array}\right);\nonumber \\
\boxdot\overline{\boxdot}C=\overline{\boxdot}\,\boxdot C= & j_{m} & =\left(\begin{array}{cc}
\rho_{m} & -\overrightarrow{j_{m}}\\
\overrightarrow{j_{m}} & \rho_{m}\end{array}\right);\nonumber \\
\boxdot\overline{\boxdot}D=\overline{\boxdot}\,\boxdot D= & j_{h} & =\left(\begin{array}{cc}
\rho_{h} & -\overrightarrow{j_{h}}\\
\overrightarrow{j_{h}} & \rho_{h}\end{array}\right)\label{eq:58}\end{eqnarray}
and

\begin{eqnarray}
\boxdot\overline{\boxdot} & = & \overline{\boxdot}\,\boxdot=(\partial_{4}^{2}+\partial_{1}^{2}+\partial_{2}^{2}+\partial_{3}^{2}).\hat{1}=(-\frac{\partial^{2}}{\partial t^{2}}+\frac{\partial^{2}}{\partial x^{2}}+\frac{\partial^{2}}{\partial y^{2}}+\frac{\partial^{2}}{\partial z^{2}}).\hat{1}.\label{eq:59}\end{eqnarray}
Hence we may write the split octionion form of unified GDM equations
in the following mannere as the connection between potential and current,

\begin{eqnarray}
\boxdot\overline{\boxdot}V=\overline{\boxdot}\,\boxdot V & =\left(\begin{array}{cc}
J_{0} & -\overrightarrow{J}\\
\overrightarrow{J} & J_{0}\end{array}\right)= & J.\label{eq:60}\end{eqnarray}
Here we may also obtain the split octonionic forms of Lorentz gauge
condition as well as the continuity equation. These are described
in terms of the inner products of two split octonions i.e. the inner
product of split octonion differential operator respectively with
potential and current of unified fields of dyons and gravito-dyons.
Similarly, Split octonion representation for field strength tensor
$Q_{\mu\nu}=V_{\mu,\nu}-V_{\nu,\mu}=F_{\mu\nu}^{EM}+F_{\mu\nu}^{GH}=(A_{\mu\nu},B_{\mu\nu},C_{\mu\nu},D_{\mu\nu})$
for the unified fields of dyons and gravito-dyons may be described
as,

\begin{eqnarray}
\mathcal{F} & = & \left(\begin{array}{cc}
\partial_{\mu}V_{\mu} & -e_{j}(\partial_{0}V_{j}+\partial_{j}V_{0}+i\varepsilon_{jkl}\partial_{k}V_{l})\\
e_{j}(\partial_{0}V_{j}+\partial_{j}V_{0}+i\varepsilon_{jkl}\partial_{k}V_{l}) & \partial_{\mu}V_{\mu}\end{array}\right)\nonumber \\
\Rightarrow & \left(\begin{array}{cc}
0 & -e_{j}\Psi_{j}\\
e_{j}\Psi_{j} & 0\end{array}\right) & \Leftrightarrow\Psi\label{eq:61}\end{eqnarray}
where $(j,k,l=1,2,3;$$\mu,\nu=0,1,2,3)$ and $i=\sqrt{-1}$ while
the field strengths $A_{\mu\nu},B_{\mu\nu},C_{\mu\nu},D_{\mu\nu}$respectively
associated with individual electric, magnetic, gravitational and Heavisidian
charges (masses) reduce to the following split octonionic forms,

\begin{eqnarray}
\mathcal{F}^{e} & = & \left(\begin{array}{cc}
\partial_{\mu}A_{\mu} & -e_{j}(\partial_{0}A_{j}+\partial_{j}A_{0}+i\varepsilon_{jkl}\partial_{k}A_{l})\\
e_{j}(\partial_{0}A_{j}+\partial_{j}A_{0}+i\varepsilon_{jkl}\partial_{k}A_{l}) & \partial_{\mu}A_{\mu}\end{array}\right)\Leftrightarrow\Psi^{e};\nonumber \\
\mathcal{F}^{g} & = & \left(\begin{array}{cc}
\partial_{\mu}B_{\mu} & -e_{j}(\partial_{0}B_{j}+\partial_{j}B_{0}+i\varepsilon_{jkl}\partial_{k}B_{l})\\
e_{j}(\partial_{0}B_{j}+\partial_{j}B_{0}+i\varepsilon_{jkl}\partial_{k}B_{l}) & \partial_{\mu}B_{\mu}\end{array}\right)\Leftrightarrow\Psi^{g};\nonumber \\
\mathcal{F}^{m} & = & \left(\begin{array}{cc}
\partial_{\mu}C_{\mu} & -e_{j}(\partial_{0}C_{j}+\partial_{j}C_{0}+i\varepsilon_{jkl}\partial_{k}C_{l})\\
e_{j}(\partial_{0}C_{j}+\partial_{j}C_{0}+i\varepsilon_{jkl}\partial_{k}C_{l}) & \partial_{\mu}C_{\mu}\end{array}\right)\Leftrightarrow\Psi^{m};\nonumber \\
\mathcal{F}^{h} & = & \left(\begin{array}{cc}
\partial_{\mu}D_{\mu} & -e_{j}(\partial_{0}D_{j}+\partial_{j}D_{0}+i\varepsilon_{jkl}\partial_{k}D_{l})\\
e_{j}(\partial_{0}D_{j}+\partial_{j}D_{0}+i\varepsilon_{jkl}\partial_{k}D_{l}) & \partial_{\mu}D_{\mu}\end{array}\right)\Leftrightarrow\Psi^{h}.\label{eq:62}\end{eqnarray}
Similarly we may write the split octonion form of the generalized
field strengths $F_{\mu\nu}^{EM}=V_{\mu,\nu}^{EM}-V_{\nu,\mu}^{EM}$of
generalized electromagnetic fields of dyons and those $F_{\mu\nu}^{GH}=V_{\mu,\nu}^{GH}-V_{\nu,\mu}^{GH}$for
the generalized gravito-Heavisidian fields of gravito-dyons in the
following manner,

\begin{eqnarray}
\mathcal{F}^{EM} & = & \left(\begin{array}{cc}
\partial_{\mu}V_{\mu}^{EM} & -e_{j}(\partial_{0}V_{j}^{EM}+\partial_{j}V_{0}^{EM}+i\varepsilon_{jkl}\partial_{k}V_{l}^{EM})\\
e_{j}(\partial_{0}V_{j}^{EM}+\partial_{j}V_{0}^{EM}+i\varepsilon_{jkl}\partial_{k}V_{l}^{EM}) & \partial_{\mu}V_{\mu}^{EM}\end{array}\right)\Leftrightarrow\Psi^{EM};\nonumber \\
\mathcal{F}^{GH} & = & \left(\begin{array}{cc}
\partial_{\mu}V_{\mu}^{GH} & -e_{j}(\partial_{0}V_{j}^{Gh}+\partial_{j}V_{0}^{GH}+i\varepsilon_{jkl}\partial_{k}V_{l}^{GH})\\
e_{j}(\partial_{0}V_{j}^{GH}+\partial_{j}V_{0}^{GH}+i\varepsilon_{jkl}\partial_{k}V_{l}^{GH}) & \partial_{\mu}V_{\mu}^{Gh}\end{array}\right)\Leftrightarrow\Psi^{GH}.\label{eq:63}\end{eqnarray}
Unified split octonion valued field tensor $Q_{\mu\nu}=V_{\mu,\nu}-V_{\nu,\mu}=F_{\mu\nu}^{EM}+F_{\mu\nu}^{GH}=(A_{\mu\nu},B_{\mu\nu},C_{\mu\nu},D_{\mu\nu})$
is self-dual and is also invariant under octonion transformations.
The components of split octonion field tensor are the components of
split octonion vector field $\Psi$ given by equation (\ref{eq:61}).
Unified split octonion valued current and split octonion field tensor
lead to the unified GDM field equation in the following manner;

\begin{eqnarray}
Q_{\mu\nu,\nu} & =J_{\mu}\Leftrightarrow & \overline{\boxdot}\,\boxdot V=\left(\begin{array}{cc}
J_{0} & -\overrightarrow{J}\\
\overrightarrow{J} & J_{0}\end{array}\right)=J\label{eq:64}\end{eqnarray}
which , on using equation (\ref{eq:56}) , is equivalent to the following
split octonion form of field equation

\begin{eqnarray}
\overline{\boxdot}\,\Psi & = & J.\label{eq:65}\end{eqnarray}
 Equation (\ref{eq:65}) is the thus represents the split octonion
formulation of GDM field equations for the unified fields of dyons
and gravito-dyons.

\section{Conclusion}

~~~~~~~~The foregoing analysis describes the combined dynamics
of dual invariant unified electromagnetic and gravito-Heavisidian
fields with the simultaneous existence of electric, magnetic, Gravitational
and Heavisidian charges (masses). Though the existence of magnetic
and Heavisidian charges is not confirmed, but sound theoretical investigations
are in favour of their existence leading to the deeper understanding
of fundamental interactions and constituents of matter. From the above
analysis it may also be concluded that besides the potential importance
of monopoles as intrinsic part of current grand unified theories,
monopoles and dyons may provide even more ambitious model to purport
the unification of gravitation with strong and electro weak forces.
The unified quaternion representation of charges in split octonion
basis shows that the dynamics of electric charge is described by the
Abelian $U\,(1)$ gauge structure while the dynamics of other charges
have the direct link with $SU(2)$ non Abelian gauge theories leading
to their extended structure. Here we have tried to accommodate a new
possibility of unification of fundamental interactions in terms of
split octonion basis elements where the advanced algebra of octonion
is capable to deal the higher dimensional structure of the theory
in order to explain the curvature in general relativity at one end
and the role of monopoles and dyons in super-symmetry, super-gravity
and super strings at the other end. The unified theory presented here
hence reproduces the dynamics of electric charge in the absence of
other charges. It also reproduces the theory of dyons in the absence
of gravito dyons or vice versa. The split octonion formalism may easily
be described for the classical and quantum theories fields associated
with one, two and four charges while the quantization condition leading
to interaction terms describe the combination of dynamics of charges
associated with the split octonion basis elements.

\textbf{Acknowledgment}- The work is supported by Uttarakhand Council
of Science and Technology, Dehradun. One of us OPSN is thankful to
Chinese Academy of Sciences and Third world Academy of Sciences for
awarding him CAS-TWAS visiting scholar fellowship to pursue a research
program in China. He is also grateful to Professor Tianjun Li for
his hospitality at Institute of Theoretical Physics, Beijing, China.

\end{document}